\documentclass[aps,pre,reprint,longbibliography]{revtex4-1}

\usepackage{amssymb}
\usepackage{amsmath}
\usepackage{cases}
\usepackage{graphicx,color}
\usepackage{hyperref}
\definecolor{r}{rgb}{1,0,0}   
\definecolor{g}{rgb}{0,1,0}   
\definecolor{b}{rgb}{0,0,1}

\begin{document}


\title{The Equation of Motion for Taut-Line Buzzers}



\author{Alexander J. Gerra$^{1,2\dagger}$, Courtney C. Jones$^{1,3\dagger}$, Sam Dillavou$^1$, Jesse M. Hanlan$^1$, Julia Radzio$^4$, Paulo E. Arratia$^4$, Douglas J. Durian$^{1,4}$}

\affiliation{$^1$Department of Physics and Astronomy, University of Pennsylvania, Philadelphia, PA 19104}
\affiliation{$^2$Moravian University, Bethlehem, PA 18018}
\affiliation{$^3$University of Maryland - Baltimore County, Baltimore, MD 21250}
\affiliation{$^4$Department of Mechanical Engineering and Applied Mechanics, University of Pennsylvania, Philadelphia, PA 19104}









\altaffiliation{$^\dagger$Equal first authors}


\date{\today}

\begin{abstract}
Equations of motion are developed for the oscillatory rotation of a disk suspended between twisted strings kept under tension by a hanging mass, to which additional forces may be applied. In the absence of forcing, damped harmonic oscillations are observed to decay with an exponential time envelope for two different string types. This is consistent with damping caused by string viscosity, rather than air turbulence, and may be quantified in terms of a quality factor. To test the proposed equation of motion and model for viscous damping within the string, we measure both the natural oscillation frequency and the quality factor for widely varied values of string length, string radius, disk moment of inertia, and hanging mass. The data are found to scale in good accord with predictions. A variation where rotational kinetic energy is converted back and forth to spring potential energy is also discussed.
\end{abstract}


\maketitle



\section{Introduction}

The ``buzzer" or ``whirligig" or ``button-on-a-string" is an age-old toy made by threading a loop of string through two holes of a disk or other symmetric solid \cite{Schlichting2010}.  By twisting up the disk then pulling on opposite ends of the loop, the disk can be set into fast rotational motion such that the string re-twists in the opposite direction and can be pulled on again, repeatedly. In each cycle, the disk starts at rest with the strings slack and fully twisted; as the strings are pulled, the disk accelerates; it reaches top speed when the strings become untwisted; then pulling ceases, the strings twist up in the opposite direction, and the disk soon comes to rest with the strings slack and maximally twisted.  The physical principles governing these rich dynamics have been previously explored~\cite{Schlichting2010, Bhamla2017}, and are important in light of potential uses as low-cost hand-powered centrifuges for medical diagnostics~\cite{Bhamla2017, Bhamla2019, Chen2020, Mukai2023} and in electricity generation and energy harvesting~\cite{Zhao2017, Tang2018, Tan2020, TangAPL2020}.  However, buzzer dynamics is difficult to analyze. Ref.~\cite{Schlichting2010} states that ``Despite its simple makeup, the quantitative analysis of the buzzer's underlying physical mechanism turns out to be very complex. Unsurprisingly, a physical model appropriate for high-school and college physics does not exist."  This is underscored in Ref.~\cite{Bhamla2017}, where analysis of high-speed behavior requires accounting for supercoiling and other effects regarding how the two strands in the loop twist together near the disk and near the hand.

\begin{figure}[th!]
\includegraphics[width=3.00in]{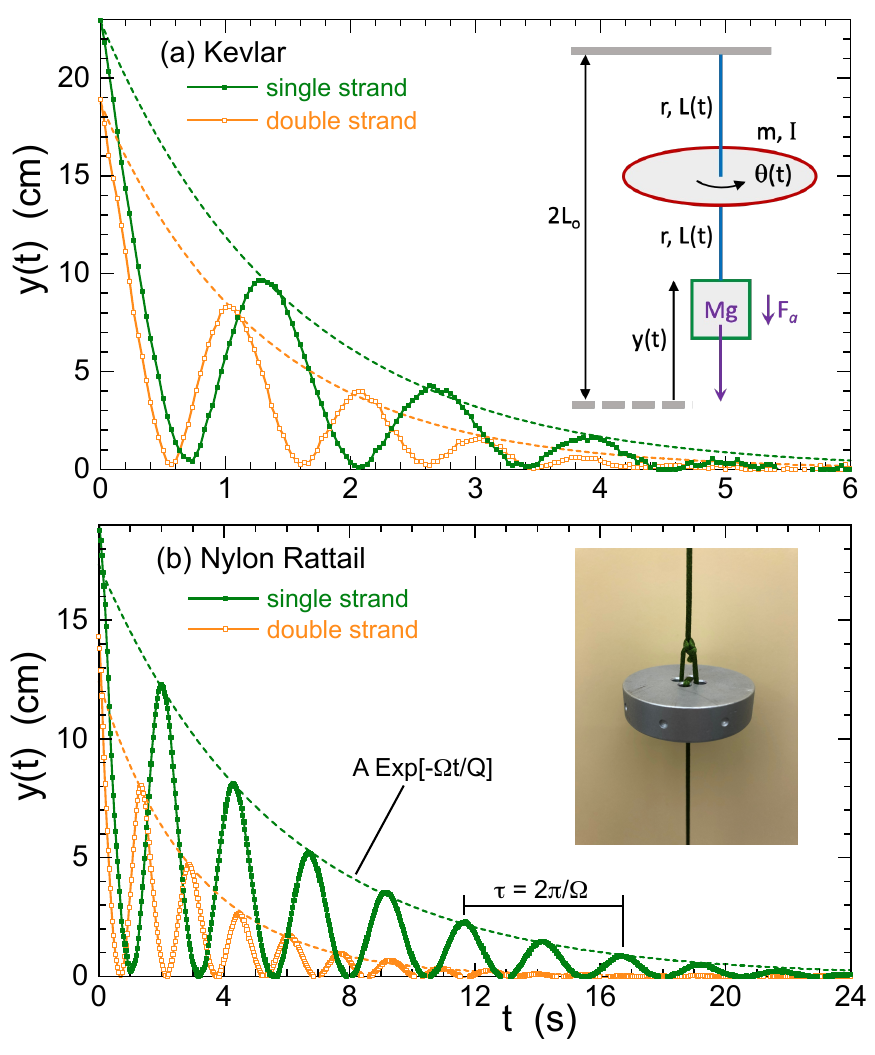}
\caption{Height versus time for ``taut-line buzzers" (top inset) made with single- and double strands of (a) Kevlar and (b) Nylon rattail string, with nominal geometric diameters of 2.1 and 2.0~mm respectively. The untwisted string lengths are all $L_o=40$~cm.  The hanging mass is $M=2175$~g. The buzzer is an Aluminum disk of radius $R=2.54$~cm and height $H=1.27$~cm (bottom inset). It was twisted by hand until the strings were about to supercoil, released at time $t=0$, and allowed to respond without any applied force ($F_a=0$). The frequency $\Omega$ and quality factor $Q$ for the damped harmonic oscillations are deduced from the time between every-other peak and the exponential decay envelope.}
\label{fig_DampedBuzzers}
\end{figure}

Here we introduce a ``taut-line buzzer'' constructed from a disk between strings held under tension by a hanging mass, as depicted in the inset of Fig.~\ref{fig_DampedBuzzers}a.  By contrast with the usual hand-pulled buzzer, we find that this setup is amenable to analysis as a simple harmonic oscillator. As such, a key difference is that the rise of the hanging mass stores gravitational potential energy, which is converted back to rotational kinetic energy of the disk as the mass falls. Therefore only a fraction of energy is lost per cycle, whereas it is entirely lost for the usual hand-pulled buzzer. This makes the taut-line buzzer easier to drive, by application of a periodic force to the hanging mass, and could enable higher and more-controlled rotation speeds~\cite{TautLineDriven}.

To begin, we develop an equation of rotational motion by considering the restoring and damping torques applied to the disk by the strings as well as from dissipation in the surrounding air. By observing the decay of oscillations, we find that it behaves as a simple harmonic oscillator with viscous damping internal to the strings. We predict the natural oscillation frequency and quality factor, and find good agreement with experiments in which system parameters are widely varied. With these insights, we hope to enable a hand-powered centrifuge that would be both easier to operate and quicker to separate components of blood for medical diagnostics~\cite{TautLineDriven}.

\section{Equation of Motion}

In a traditional hand-pulled buzzer, torque is applied to a disk by periodically tugging on a loop of twisted strings. All the rotational kinetic energy imparted to the spinning disk is then lost each cycle as the disk decelerates and twists up the strings in the opposite direction. We seek an alternative design where the kinetic energy of the spinning disk is converted to some form of potential energy, which can then be recovered and converted back to rotational kinetic energy. The proposed device shown schematically in the top inset of Fig.~\ref{fig_DampedBuzzers} achieves this via a hanging mass $M$ that keeps the string taut and rises upwards to store gravitational potential energy when the strings twist up and the disk decelerates.  Hence we call it a ``taut-line buzzer." The disk is characterized by mass $m$, moment of inertia $I$, and its angular position versus time $\theta(t)$. If the buzzer is a disk of density $\rho$, radius $R$, and height $H$, then $m=\rho \pi R^2H$ and $I=mR^2/2$. For simplicity we use a symmetrical design where the untwisted string lengths are $L_o$ both above and below.

When twisted by angle $\theta$, strings and bundles of strings of initial length $L_o$ contract according to an effective twist radius $r$ according to
\begin{equation}
    L = \sqrt{L_o^2 - (r\theta)^2}\approx L_o - \frac{(r\theta)^2}{2Lo},
\label{contraction}
\end{equation}
where the approximation is for small contraction, $r\theta\ll L_o$~\cite{Shoham2005, Guzek2012, Palli2013, Gaponov2014, Hanlan2023}. This assumes an ideal string or bundle that is inextensible, perfectly flexible with zero bending or shear moduli, and of constant radius. Refs.~\cite{Schlichting2010, Bhamla2017} present geometrical arguments for Eq.~(\ref{contraction}) and for the torque $\mathcal{T}$ applied to the disk. We find the latter to be more readily understood by an energetic argument.  In particular, when under tension $T$, the work to impose a small twist $d\theta$ is $\mathcal{T} d\theta$ and this must equal $TdL$ where $dL$ is the length contraction. Hence, the torque along the string is set by $T$ and the form of $L(\theta)$ as 
\begin{equation}
    \mathcal{T} = T \frac{dL}{d\theta}.
\label{torquegen}
\end{equation}
We are unaware of this general formula in prior literature. Combining with Eq.~(\ref{contraction}) gives a linear spring-like restoring torque for small contraction:
\begin{equation}
    \mathcal{T} = -T\frac{r^2}{L(\theta)}\theta \approx -T\frac{r^2}{L_o}\theta
\label{torque}
\end{equation}
We restrict attention to this limit for developing the equation of rotational motion. It would be straightforward to include the full nonlinearity of the restoring torque, if $L(\theta)$ were known; however, analysis must then be numerical. Since Eqs.~(\ref{torquegen}-\ref{torque}) assume that the string is inextensible and perfectly flexible, no energy is stored elastically. By contrast, for a twisted pair of elastic rods with fixed endpoints, elastic energy was recently found to be stored primarily in stretch rather than bend or rod-rod contact and analyzed in terms of a shear modulus~\cite{Kudrolli2023}.

Let us now consider a driving force $F_a$ applied to the hanging mass (positive if downward). If the vertical acceleration of all the masses can be neglected, then $Mg+F_a$ is the tension in the lower string and $(M+m)g+F_a$ is the tension in the upper string.  The general equation of rotational motion of the disk is then
\begin{equation}
    I\ddot\theta = -\left[(2M+m)g+2F_a\right]\frac{r^2}{L_o}\theta - \beta_1\dot\theta - \beta_2\dot\theta^2
\label{eqnofmotion}
\end{equation}
where the first term is the linear restoring torque from Eq.~(\ref{torque}), the $\beta_1$ term represents a viscous drag from air and/or the string, and the $\beta_2$ term represents an inertial drag from air turbulence or perhaps in some cases from sound emission (hence the name ``buzzer''). Our taut-line buzzers are, however, silent.

As an alternative kinematic variable, the equation of motion could be written in terms of the height $y$ of the hanging mass above its rest position. This height is related to twist angle via the contraction of both strings as
\begin{equation}
	y = 2(L_o-L) \approx (r\theta)^2/L_o.
\label{massheight}
\end{equation}
This can be easier to measure than $\theta$. Importantly for physical understanding, $y$ gives the gravitational potential energy of the system as $U = (M+m/2)gy$, which increases as the hanging mass rises and the disk decelerates and which can be converted back to rotational kinetic energy of the disk as the hanging mass falls.

\subsection{Sources of dissipation}

To design an easy-to-drive taut-line buzzer, energy dissipation mechanisms must be understood and minimized. The order of magnitude for various possible sources of drag may be estimated as follow.  First, the importance of viscous versus inertial contributions to air drag on the disk is set by the Reynolds number, $\textrm{Re}=\rho_a \dot\theta R^2/\eta_a$, where $\rho_a=0.001255$~g/cc and $\eta_a=0.000182$~g/(cm~s) are the mass density and viscosity of air.  Even for $R=5$~cm and a very slow angular speed of $\dot\theta=60$~RPM, the Reynolds number is already large, $\textrm{Re}\approx 1000$; therefore, for any interesting buzzer or hand-powered centrifuge, Re will be even larger and therefore viscous air drag is unimportant. For inertial air drag, the order of magnitude for the retarding torque must scale as pressure times area times lever arm, \textit{i.e.} $\beta_2 \dot\theta^2 \approx [\rho_a(\dot\theta R)^2] \times (R^2+RH) \times R$, ignoring numerical factors.  This gives
\begin{equation}
    \beta_2 \approx \rho_a(R+H)R^4
\label{beta2}
\end{equation}
with an unknown numerical prefactor likely to depend on Re and be small (the spinning disk displaces little air). A similar expression for turbulent air drag was given in Ref.~\cite{Bhamla2017}, and used for numerical modeling of hand-pulled buzzer dynamics.

Other possibilities involving inertia, and hence the $\beta_2$ term, arise from non-ideal effects. For example the center of mass of the disk could be off-axis and cause transverse shaking.  Or if the buzzer shape is not a solid of revolution, it will generate significantly more turbulence and probably transverse shaking. Or transverse vibrational modes of the string could be excited, resonant with the disk oscillations or not, which cause the string to cut through the air as it spins. In fact we sometimes observe this phenomenon with hand-pulled buzzers, which creates a loud whooshing noise. If the buzzer is constructed from a loop of string, we also sometimes observe the two strands to fly apart when untwisted, which corresponds to highest speed.  This too ``whooshes" as the strings cut through the air before coming together and retwisting. All such motions are complicated and beyond our scope to model, but clearly dissipate energy.

Another source of dissipation, which can be readily modeled, arises from the viscosity $\eta_s$ associated with shear deformation throughout the volume of the string. For this, the retarding torque scales as viscosity times shear strain-rate times area times lever arm, \textit{i.e.} $\beta_{1} \dot\theta \approx \eta_s \times [(\dot\theta r)/L_o] \times r^2 \times r$, which gives
\begin{equation}
    \beta_{1} \approx \eta_s r^4/L_o
\label{beta1}
\end{equation}
with an unknown numerical prefactor. Dissipation within strings was mentioned in Ref.~\cite{Zhao2017}, but to our knowledge has never previously been modeled. One could also consider dissipation due to fiber-fiber sliding friction in a braided string. Presumably this should be rate-independent and proportional to string tension, which is inconsistent with observations below.  Therefore, inertial air drag and viscous string damping are the two most likely sources of dissipation.

To gauge the relative importance of inertial air drag to viscous string damping, we define a dimensionless number
\begin{equation}
    B = \frac{\beta_2 \dot\theta^2}{\beta_1 \dot\theta} \approx \frac{\rho_a L(R+H)R^4}{\eta_s r^4}\dot\theta
\label{buzzernumber}
\end{equation}
in analogy with the Reynolds number. At small $B$ string viscosity dominates; at large $B$, \textit{i.e.} at high enough angular speed, inertial air drag dominates. For a given choice of string and disk dimensions, one needs to know a typical or maximum angular speed of the disk as well as the viscosity of the string.  This requires experiment -- as reported after the following subsection.

\subsection{Damped oscillations}

In the absence of forcing, $F_a=0$, the equation of rotational motion may be rewritten
\begin{equation}
    \ddot\theta = -\Omega_o^2 \theta - \frac{1}{Q}\Omega_o\dot\theta - \frac{1}{Q_2}\dot\theta^2
\label{EOM3}
\end{equation}
where the natural oscillation frequency, the conventional $Q$ factor for viscous damping \cite{Marion}, and a kind of quality factor for inertial drag, are respectively defined as
\begin{eqnarray}
    \Omega_o &=& \sqrt{\frac{(2M+m)gr^2}{IL_o}} \label{Omega} \\
    Q &=& \Omega_o I/\beta_1 \approx \frac{\sqrt{(2M+m)gIL_o}}{\eta_s r^3} \label{Qfactor} \\
    Q_2 &=& I/\beta_2 \label{Q2def}
\end{eqnarray}
A similar expression for the oscillation frequency is given in Section~3.5 of Supplementary Information in Ref.~\cite{Bhamla2017}, where twice the peak amplitude of the hand-applied force replaces $(2M+m)g$. Here the $Q$ factor is further expressed using the definition of $\Omega_o$ and the Eq.~(\ref{beta1}) estimation of $\beta_1$ for the case of string viscosity. If the damping is viscous, then the total mechanical energy decays as $E(t) \propto \exp(-\Omega_o t/Q)$ and the fractional energy loss per cycle is given by $Q=2\pi E/\Delta E$. If the viscous damping is small, as turns out to be the case for our taut-line buzzers (next), then the predicted oscillation frequency $\Omega=\Omega_o\sqrt{1-1/(4Q^2)}$ is negligibly decreased from the natural oscillation frequency.

\section{Experimental Tests}

\subsection{Source of damping}

To begin we first measure damped oscillations for four different taut-line buzzers made with an Aluminum disk of radius $R=2.54$~cm and height $H=1.27$~cm (Fig.~\ref{fig_DampedBuzzers}b inset), and a hanging mass $M=2175$~g fashioned from a 2~liter plastic bottle partially filled with steel bearing balls. The untwisted string lengths are all $L_o=40$~cm.  The string materials are Kevlar (polyparaphenylene terephthalamide; Emma Kites) and Nylon rattail (polyamide; Konmay) with respective manufacturer-quoted diameters of $r=2.1$~mm and 2.0~mm. Both are composed of braided fibers of a liquid-crystal polymer. To initiate damped oscillations, the disk is twisted by hand until the strings are about to supercoil.  This maximum allowed twist angle is roughly $\theta_{s} \approx(0.8-0.9)L_o/r$~\cite{Hanlan2023}.  Then the disk is released and allowed to freely gain angular speed as the string untwists and the hanging mass falls. The moment of inertia of the hanging mass is much greater than that of the disk, but it is not infinite.  Therefore, some small nwanted rotational motion of the hanging mass is induced; see Eqs.~(\ref{EOM1},\ref{EOM2}). Such motion is minimized by attaching three slightly-slack 1~m strings to the bottom of the hanging mass and anchoring them to the surroundings. Indistinguishable results are obtained with two 10~cm anchor lines held by hand and moved up and down in parallel with the mass. A range finder (Vernier, Motion Detector 2 and Logger Pro) is placed on the floor under the hanging mass in order to measure its time-dependent height, $y(t)$. Recall that this relates to gravitational potential energy via $U=(M+m/2)gy$.

Results for $y(t)$ versus $t$ are plotted in Fig.~\ref{fig_DampedBuzzers} for single- and double strands of both string materials. For each there are several oscillations that decay with an exponential envelope as $y(t) = y_o \cos^2(\Omega t)\exp(-\Omega_o t/Q)$. Note that the period $\tau=2\pi/\Omega$ is the time between every other peak in $y(t)$, because the strings are fully twisted in opposite directions at adjacent peaks. Such oscillatory behavior is consistent with lightly-damped simple harmonic motion of $\theta(t)$ according to the equation of rotational motion (\ref{eqnofmotion}) without the $\beta_2$ inertial drag term. Since several oscillation cycles are apparent before motion ceases, the $Q$ values are of order 10.  We conclude that string viscosity must be the dominant source of damping. This is strongly supported by the observation that $Q$ varies with string material when all else is held fixed.  Based on the number of oscillation periods in the decays seen in Fig.~\ref{fig_DampedBuzzers}, string viscosity is noticeably larger for Kevlar than for Nylon rattail. Indeed, the Nylon rattail feels more slippery while Kevlar is known to be self-abrading under dynamic loading.  Furthermore, we find it considerably easier to drive steady oscillation of both hand-pulled and taut-line buzzers made with Nylon rattail than with Kevlar.

\begin{table}
\caption{\label{rattails}%
Nominal and measured twist radii for different sizes of braided Nylon rattail cord (Konmay), purchased in different colors.}
\begin{ruledtabular}
\begin{tabular}{lcc}
String & $r_n$ (mm) & $r_t$ (mm) \\
\colrule
red & 0.50 & $0.60\pm0.02$ \\
gold & 0.75 & $0.83\pm0.02$ \\
green & 1.0 & $0.95\pm0.02$ \\
blue & 1.25 & $1.09\pm0.03$ \\
black & 1.50 & $1.91\pm0.03$ \\
\end{tabular}
\end{ruledtabular}
\end{table}

\begin{table}
\caption{\label{buzzers}%
Radius, height, mass, and moment of inertia for buzzer disks machined from Aluminum plate.}
\begin{ruledtabular}
\begin{tabular}{ccccc}
Buzzer & $R$ (in) & $H$ (in) & $m$ (g) & $I$ (g$\cdot$cm$^2$) \\
\colrule
A & 1/2 & 3/8 & 10.9 & 8.79 \\
B & 1 & 1/4 & 33.2 & 107. \\
C & 1 & 3/8 & 49.9 &  161. \\
D & 1 & 1/2 & 68.9 &  222. \\
E & 1 & 1 & 133.6 & 431. \\
F & 1 & 2 & 267.8 & 864. \\
G & 2 & 3/8 & 215.8 & 2780. \\
\end{tabular}
\end{ruledtabular}
\end{table}

\begin{figure}[t]
\includegraphics[width=3.00in]{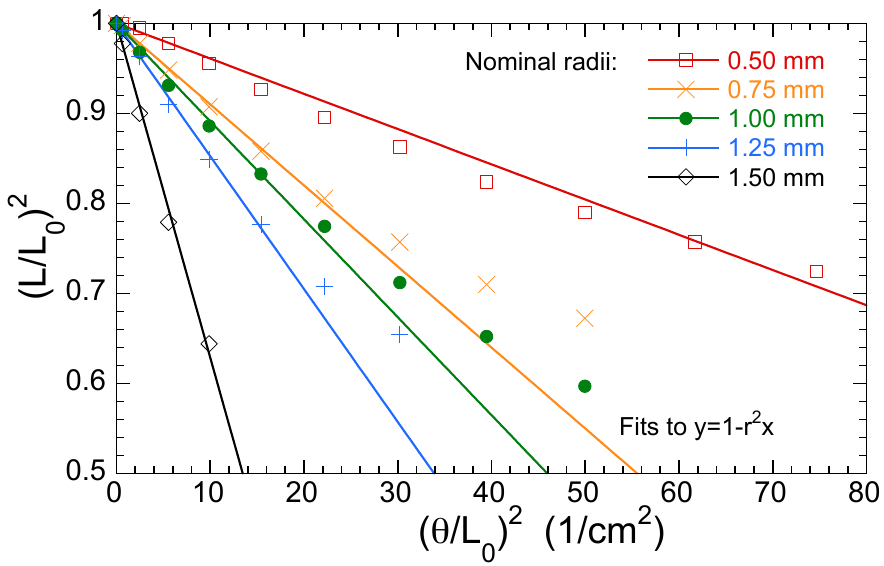}
\caption{Contracted length $L$ of nylon rattail strings of various radii versus twist angle $\theta$, divided by the untwisted string length $L_o=40$~cm and squared. The solid lines are fits to Eq.~(\ref{contraction}), rewritten as $(L/L_o)^2 = 1 - (r\theta/L_o)^2$, for the effective twist radii $r$; results are given in Table~\ref{rattails}.}
\label{fig_contraction}
\end{figure}

\subsection{Scaling of frequency and quality factor}

To quantitatively test the Eq.~(\ref{Omega},\ref{Qfactor}) predictions for $\Omega_o$ and $Q$, we make similar measurements of damped oscillations using the same procedures as above, but now versus widely varied system parameters.  For this we utilize single strands of braided Nylon rattail cord with radii listed in Table~\ref{rattails} and lengths $L_o=\{20,30,40,\ldots,90\}$~cm. The manufacturer's reported (nominal) radii correspond well with visual inspection but not necessarily with the twist radii defined by Eq.~(\ref{contraction}) because of their braid and fiber structure. Therefore we measure contraction versus twist angle for $L_o=40$~cm untwisted lengths of each size.  Results are plotted in Fig.~\ref{fig_contraction} as $(L/L_o)^2$ versus $(\theta/L_o)^2$, along with fits to a line for the twist radius.  These fits are reasonably good, but there are noticeable systematic errors (the origin of which is explored in Ref.~\cite{Hanlan2023}).  The resulting twist radii in Table~\ref{rattails} differ somewhat from the nominal radii -- some are larger and some smaller.  Buzzers disks are machined from Aluminum plates with radii, heights, masses, and moments of inertia listed in Table~\ref{buzzers}. Each has four concentric 3/16 inch diameter holes located 3/16 inches from the axis, as well as small dimples machined into the sides for potential use in high-speed video imaging of angular position. The hanging masses are $M=\{1, 2, 2.175, 3, 4, 5\}$~kg. Damped oscillations are measured with all parameters --except the one being varied-- fixed at $L=40$~cm, $r=1$~mm, buzzer~D, and $M=2.175$~kg; for trials versus hanging mass, $M=2.175$~kg is not used.

\begin{figure}[t]
\includegraphics[width=3.00in]{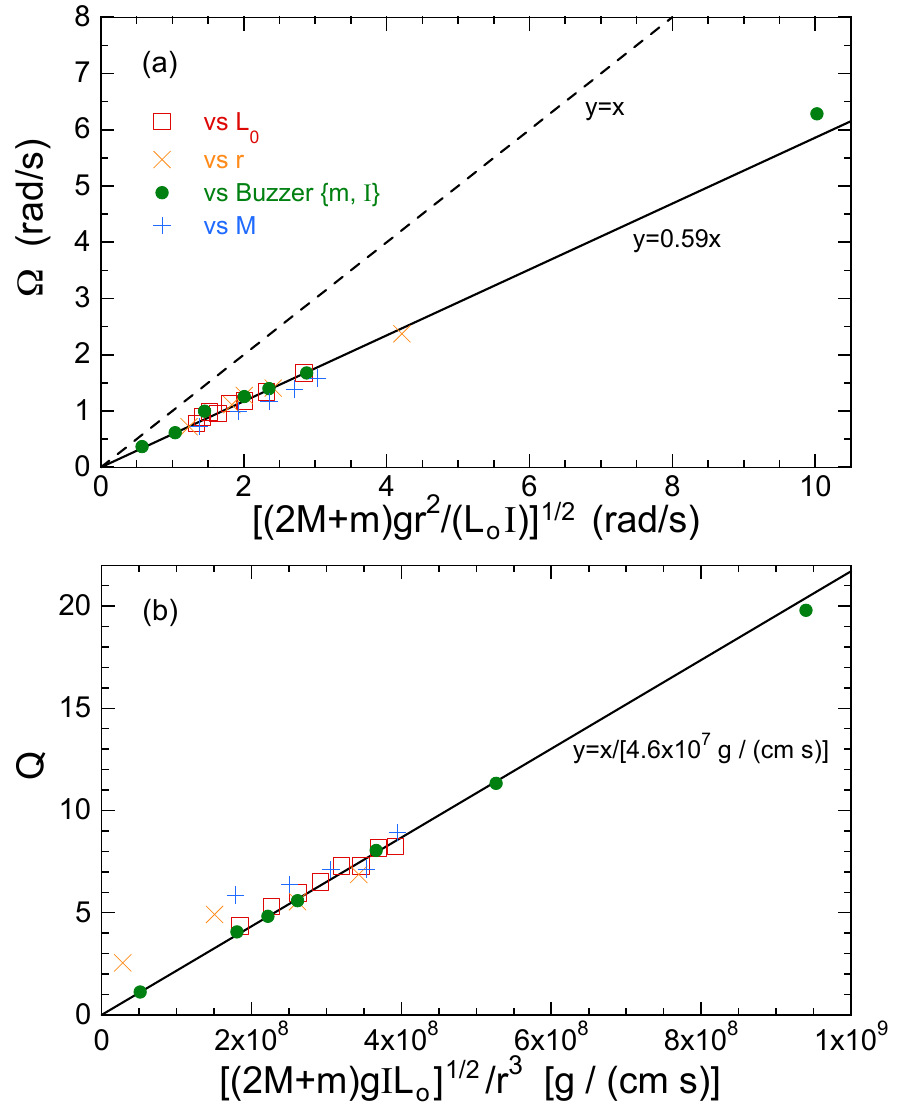}
\caption{(a) Natural frequency and (b) quality factor data for damped oscillations versus prediction. Trials are made by fixing the untwisted string length $L_o$, string radius $r$, buzzer mass $m$ and moment of inertia $I$, and the hanging mass $M$ to median values and then varying one at a time across a range of values given in the text.}
\label{fig_OmegaQ}
\end{figure}

From the $y(t)$ versus $t$ range finder data we extract the period $\tau$ from the time between every second peak, giving frequency $\Omega = 2\pi/\tau$. And by fitting the peak heights to $y_o \exp(-\Omega_o t/Q)$, we extract the $Q$ factor. These are generally large enough that the distinction between $\Omega_o$ and $\Omega=\Omega_o\sqrt{1-1/(4Q^2)}$ was neglected compared to experimental uncertainties.  Results for oscillation frequency are plotted in Fig.~\ref{fig_OmegaQ}a versus the predicted combination of system parameters given in Eq.~(\ref{Omega}). The data collapse nicely to the line $y=0.59x$; therefore, the observed frequencies trend with \emph{all} system parameters in accord with expectation but are about 2/3 as fast. Perhaps this could be explained by nonlinearity of the restoring force for large amplitudes, evidenced by the systematic deviation between data and fit for $r$ seen in Fig.~\ref{fig_contraction} (where the slopes and hence effective spring constants are reduced for large twist angles, as seen earlier \cite{Hanlan2023}).  Results for the quality factor are plotted in Fig.~\ref{fig_OmegaQ}b versus the predicted combination of system parameters given in Eq.~(\ref{Qfactor}). The data collapse nicely to the line $y=x/[4.6\times 10^7$~g/(cm$\cdot$s)]; therefore, the trends versus known system parameters are in accord with expectation and the viscosity of all Nylon rattail strings under all tensions is of order $\eta_s \approx 10^7$~g/(cm$\cdot$s) = 1~MPa$\cdot$s.

The string viscosity is comparable to high molecular weight ($M_w$) molten polymers such as polyisobutene and poly-demethyloxane \cite{Kumar_MeltsReview_1980} and other polymer blends \cite{Bird87, DEALY2006}, which may seem high. For comparison, the viscosity of molten Nylon is relatively constant at strain rates lower than 1 s$^{-1}$ (depending on $M_w$ and processing conditions)\cite{Khanna_PES_1996, Seo_ACS_2018} and increases nonlinearly (up to 10$^5$ Pa$\cdot$s) as temperature is lowered toward the melting point (260--270~C)\cite{Kumar_MeltsReview_1980, Pezzin_1964, dePablo_JPS_2022}. But of course the nylon is solid, so the following numbers may be relevant. First, the maximum possible strain is set by the maximum allowed twist angle $\theta_s \approx L_o/r$ \cite{Hanlan2023} as $\gamma_s = (\theta_s r) / L_o \approx 1$. This is not small compared to typical yield strains; visually, however, the Nylon strings all remain undamaged, even after repeated use (by contrast with the Dyneema strings in Ref.~\cite{TautLineDriven}). So plastic flow within individual fibers does not seem to be the source of viscous dissipation. Sliding friction from the rubbing of fibers within the braid must occur, but can perhaps be ruled out since it ought to be rate-independent and depend on the tensile stress $\approx M g / r^2$, which is widely varied with no measured effect on $\eta_s$. Altogether, a reasonable hypothesis is that viscous dissipation is due to breakage and reformation of hydrogen bonds between polyamide chains, which are aligned along the length of the fibers~\cite{dePablo_JPS_2022, Xu2020}.  In terms of rate, the period of oscillations is typically longer than one second; therefore, the typical maximum strain rate is less than $1~{\textrm s}^{-1}$, which is in the quasi-static regime -- at least for compression of solid Nylon~\cite{Xu2020}.

\section{Variations}

Before closing, we generalize to a taut-line buzzer consisting of a mass $m_1$ with moment of inertia $I_1$ suspended by a string of untwisted length $L_1$ and radius $r_1$, below which a second mass $m_2$ with moment of inertia $I_2$ is suspended by a string of untwisted length $L_2$ and radius $r_2$. This second mass is subject to an applied force $F_a$ (positive if downward), and is now also allowed to rotate. The angular coordinates of the two masses are $\theta_1$ and $\theta_2$, respectively, measured in the same direction. Thus, the lower string is twisted opposite the top string by angle $\theta_1-\theta_2$. Ignoring dissipation and forcing for now, the total torques exerted on the two masses give the following coupled equations of motion:
\begin{eqnarray}
	I_1\ddot\theta_1 &=& -T_1\frac{r_1^2}{L_1}\theta_1 -T_2\frac{r_2^2}{L_2}(\theta_1-\theta_2)
        \label{EOM1} \\
	I_2\ddot\theta_2 &=& -T_2\frac{r_2^2}{L_2}(\theta_2-\theta_1) \label{EOM2}
\end{eqnarray}
where the string tensions are $T_2=m_2g+F_a$ and $T_1=T_2+m_1g$. These equations could be analyzed for normal modes. They can also be derived from the Lagrangian $\mathcal{L}=K-U$ where $K=(1/2)[I_1\dot\theta_1^2 + I_2\dot\theta_2^2 + m_1\dot y_1^2 + m_2\dot]$, $U=m_1gy_1+m_2gy_2$, and the respective rise of the masses
\begin{eqnarray}
    y_1 &=& \frac{(r_1\theta_1)^2}{2L_1} \label{y1} \\
	y_2 &=& y_1+\frac{[r_2(\theta_2-\theta_1)]^2}{2L_2} \label{y2}
\end{eqnarray}
This alternative derivation, which by-passes torque, underscores our contention that string contraction versus twist angle is a crucial physical ingredient for understanding buzzer dynamics. The Lagrangian method also makes it straightforward to account for the vertical translational kinetic energy of the masses via inclusion of $(1/2)m_1 \dot y_1^2 + (1/2)m_2 \dot y_2^2$.  This gives cumbersome equations of motion that are nonlinear. For example, including the translation kinetic energy of the hanging mass and returning to the experimental conditions ($m_1=m$, $m_2=M$, $r_1=r_2=r$, $L_1=L_2=L$, $\theta_1=\theta$, $\theta_2=0$, $I_1=I$) gives
\begin{equation}
    \frac{d}{dt}\left[\left(I+4M\frac{r^4}{L^2}\theta^2\right)\dot\theta\right]
    = -\left[(2M+m)g - 4M\frac{r^2}{L}\dot\theta^2 \right]\frac{r^2}{L}\theta.
\label{HangingKE}
\end{equation}
This can be compared with Eq.~(\ref{eqnofmotion}) for deciding when it is valid to neglect the kinetic energy of the hanging mass. For example, for a maximum possible twist angle of $\theta_\mathrm{max}\approx L/r$ \cite{Hanlan2023} and maximum angular speed of $\dot\theta_\mathrm{max}=\theta_\mathrm{max}\Omega_o$, the two extra nonlinear terms in Eq.~(\ref{HangingKE}) can both be neglected if $4Mr^2 \ll I$ holds. For our experiments $4Mr^2$ ranges from 14 to 450~g$\cdot$cm$^2$ while $I$ ranges from 8.8 to 2800~g$\cdot$cm$^2$. Therefore, we were careful to make measurements only for small twist angles.

Lastly, we consider using a spring instead of (or in supplement to) gravity as a way to store rotational kinetic energy of the buzzer, so that it can be recovered rather than lost each cycle.  Specifically, if a spring of constant $k$ is placed in line with the strings, it will be stretched by some initial amount $s_o$ plus $y_2$ given above. The tensions in the strings then both increase by $k(s_o+y_2)$ and this modifies the equations of motion accordingly. Returning to the case of identical strings above and below the buzzer, assuming the spring force is strong compared to gravity, and including forcing plus viscous dissipation, we have the following nonlinear equation of rotational motion for the spring-loaded buzzer:
\begin{equation}
    I\ddot\theta = -2\left[k\left(s_o+\frac{r^2\theta^2}{L_o}\right)+F_a\right]\theta - \beta_1\dot\theta
    \label{SpringEOM}
\end{equation}
We have not analyzed this prediction, but we did build such a device using an archery bow, a button, and a single strand of Kevlar string. Steady driving was accomplished by placing one end of the bow on the floor and pushing down on the top end as the string twisted up. This device is related to those of Refs.~\cite{Tan2020, TangAPL2020}, without a twisted loop or significant coupling with translational inertia. High enough speeds were reached that the Kevlar became hot and started smoking. This also supports our conclusion that the dominant source of dissipation is within the strings.

\section{Conclusion}

In this paper we introduced the taut-line buzzer as a device with several potential advantages over the usual hand-pulled buzzer. From observation of damped harmonic oscillations for two string materials and for widely varying system parameters, we conclude that drag is dominated by viscous damping within the string. Viscous and turbulent air drag are both negligible by comparison.  The governing equation of rotational motion is thus Eq.~(\ref{eqnofmotion}) with $\beta_1$ given by Eq.~(\ref{beta1}) and $\beta_2$ set to zero, making it amenable to elementary analysis. This was specifically demonstrated by the scaling of the oscillation frequency and the quality factor with system parameters in Fig.~\ref{fig_OmegaQ}. In the absence of driving, the taut-line buzzer behaves as a conventional damped harmonic oscillator, where rotational kinetic energy and gravitational potential energy convert back and forth each cycle and where the energy loss can be small. For the usual hand-pulled buzzer, by contrast, all mechanical energy is totally lost and must be re-injected each cycle. A properly-designed taut-line buzzer should thus be easier to drive at higher speed and for longer duration, as needed for fun or application. This is discussed in another paper \cite{TautLineDriven}. For now, note that the equation of motion is nonlinear -- by contrast with a driven simple harmonic oscillator -- because the driving force adds to the tension and hence is multiplied by the twist angle in Eq.~(\ref{eqnofmotion}).


\begin{acknowledgments}
We thank Peter Olmsted for discussions about the molecular origins of string viscosity. This work was supported by NSF grants REU-Site/DMR-1659512 as well as MRSEC/DMR-1720530 and MRSEC/DMR-2309043. DJD
thanks CCB at the Flatiron Institute, a division of the Simons Foundation, as well as the Isaac Newton Institute for Mathematical Sciences under the program ``New Statistical Physics in Living Matter" (EPSRC grant EP/R014601/1), for support and hospitality while a portion of this research was carried out.
\end{acknowledgments}


\bibliography{BuzzerRefs}

\end{document}